\documentclass{elsart}
\usepackage{epsfig,natbib,amssymb}
\journal{Advances in Space Research}

\def\deg {\hbox{$^\circ$} }
\def\eq#1{\begin{equation} #1 \end{equation}}

\def\Rmeteor{R_{\rm m}}
\def\Vmeteor{v_{\rm m}}

\begin{document}

 \begin{frontmatter}
 \title {Thermalization of Sputtered Particles as the Source of
           Diffuse Radiation from High Altitude Meteors}

 \author{Dejan Vinkovi\'{c}}
 \address{School of Natural Sciences, 
          Institute for Advanced Study, 
          Einstein Drive, Princeton, NJ 08540, USA}
 \ead{dejan@ias.edu}

 \begin{abstract}

High altitude meteors become luminous at altitudes above $\sim$130 km,
where the standard ablation theory of meteor light production is not
applicable. The physical mechanism responsible for their glow has not
been known.  We present a model that explains their existence,
morphology and lightcurves.  The model is based on particles ejected
from the meteoroid surface through the sputtering process.  The
kinetic energy of such a sputtered particle is typically more than
1,000 times larger than the energy of particles in the surrounding
atmosphere. Thus the sputtered particle creates a cascade of
collisions in the atmosphere during thermalization. We show
analytically that this process is capable of producing enough light
for detection.  We also explain the observed relationship between the
beginning height of high altitude meteors and their maximum
brightness. In addition, meteors are modeled with a Monte Carlo code
developed specifically for this phenomenon.  Theoretical images
reproduce the observed shapes and sizes of high altitude
meteors. Their exact shape depends on the relative angle between the
meteor path and observer's line of sight.

 \end{abstract}

 \begin{keyword}
 Meteor; Sputtering; High altitude meteors; Monte Carlo; Imaging;
 \end{keyword}
 \end{frontmatter}

\section{Introduction}

A meteor is a luminous phenomenon produced by a meteoroid colliding
with the earth's atmosphere. The collision starts with atmospheric
particles impinging on the meteoroid's body, by which the surface
heats up and sublimates. The cloud of vapor particles surrounding the
meteoroid collides further with the atmosphere and becomes ionized and
excited. The meteor light is produced by deexcitations, with metals
from the meteoroid dominating the spectrum. When the meteor reaches
altitudes where the Knudsen number $Kn$ (ratio between the atmospheric
mean free path and the meteoroid size) is less than $\sim$100, vapors
completely engulf the body and shield it form direct collisions with
the atmosphere. After that the heating and evaporation of the
meteoroid is caused by hot vapor and radiation from the shock front.
During the whole process the meteoroid body is ablating, that is
losing mass either as gas or as fragments. The emission of light
exists only as a consequence of the ablation process.

The limiting altitude at which most meteors become luminous is
$\sim$120 km or less, with very few starting at $\sim$130 km
\citep[e.g.][]{Betlem, Campbell, Brown}. The atmospheric mean free
path is on the order of meters at these altitudes, which is enough to
induce intense evaporation of volatiles from large (centimeters in
size) fast meteoroids. It came as surprise, therefore, when meteors of
much higher beginning heights were detected in 1995 and 1996 by
\citet{Fujiwara} using TV cameras. They reconstructed trajectories of
two Leonid fireballs of -4$^m$ and -7$^m$ visual magnitude and found
that they started at $\sim$160 km altitude. This is about 30 km higher
than the highest meteors detected by photographic systems
\citep{Betlem}. The detection was attributed to the much higher
sensitivity of TV cameras, especially the infrared region of 900 nm.

This result was confirmed and extended further in 1998 by
\citet{Spurny2000a}, using high-sensitivity TV systems with limiting
magnitude +6.5$^m$. They detected thirteen Leonid fireballs at
altitudes above 130km. Three of them were above 180km, with the
highest at 200 km. The atmospheric mean free path is over 200 m at an
altitude of 200 km, and more than 50 m at 160 km, while the meteoroid
body has a size of no more than tens of centimeters for the brightest
detected meteors.  It is puzzling how meteors can produce light in
such conditions.  The meteoroid surface is still too cold for
sublimation or evaporation and, thus, these {\it high altitude}
meteors can not be explained by the standard theory of meteor
ablation.

\citet{Spurny2000b} described the morphological features of these
meteors.  Initially they appear as a several kilometers big diffuse
spot that rapidly transforms into a diffuse comet-like V-shape
structure with a head and tail. Sometimes structures like arcs, jets
and streamers are visible in the tail.  At altitudes of $\sim$130 km
the meteor changes its appearance into an ordinary droplet-like head
shape with trail. This also marks the moment when the standard
ablation theory becomes applicable. High altitude meteors up to 150 km
were also detected in other meteor streams: $\eta$-Aquarids, Lyrids,
and Perseids \citep{Koten}.

The decreasing size of the diffuse structure with altitude suggests
that the atmospheric mean free path is important as the light is
induced by collisions.  \citet{Spurny2000a} noted that the beginning
meteor height grows with the peak brightness of meteors. This implies
that the number of collisions increases with the meteoroid
size. Considering these properties of high altitude meteors, we
suggest that they are produced by a cascade of collisions in the
atmosphere during thermalization of particles sputtered from the
meteoroid surface. Sputtering is a process in which an energetic
particle projectile ejects atoms from the surface of a solid
material. It has been used in meteor physics only to study heating of
the surface or slowing down of micrometeoroids
\citep{Opik,Lebedinets,Bronshten,Coulson}.  The luminosity has been
associated exclusively with the phase of intense
evaporation\footnote{During the preparation of this manuscript, we
became aware of a study by Hill, Rogers \& Hawkes (accepted for
publication in {\it Earth. Moon, and Planets}) describing sputtering
as a possible source of luminosity in high altitude meteors.}, during
which the number of released atoms is increased dramatically
\citep{Bronshten}.  Unlike vaporization, sputtering can exist at high
altitudes because it does not depend much on the meteoroid surface
temperature.

It is the purpose of this work to reproduce basic properties of high
altitude meteors. We present our 3D Monte Carlo model that traces
multiple particle collisions within the atmosphere and show modeling
results in the form of theoretical images from a ground-based CCD
detector (TV camera).

\section{Sputtering from the meteoroid surface}

Meteoroids in meteor showers enter the earth's atmosphere with
geocentric velocity $\sim$15 km/s to 71 km/s. From the point of view
of the meteoroid surface the atmospheric molecules behave like
projectiles with kinetic energies of hundreds of eV. For example, an
oxygen molecule at 71 km/s has 840 eV of kinetic energy. Collision
with the surface dumps this energy into displacements of atoms in the
solid until the projectile stops. If the energy is high enough, some
atoms are ejected from the surface, and we call these {\it sputtered
particles}.  If the projectile exits the solid then we call it a {\it
scattered particle}.

The {\it sputtering yield} $Y$ is defined as the number of sputtered
particles relative to the number of projectiles. The yield depends on
the projectile energy, physical and chemical properties of the solid
target, and on the type of the sputtered atom. The experimental data
on sputtering at the energies of interest here are scarce, thus we
mostly rely on numerical models.  Since the high altitude meteors
detected so far belong to cometary material, we can assume that they
are made of a fragile (low binding energy) amorphous material.
Calculations show that sputtering yields on such targets approach
$Y$=1 for projectile energies of hundreds of eV and that the sputtered
atoms have typically 5-10\% of the impact energy \citep{Field}.  In
the case of silicates typically present in interstellar grains, the
yield can be one or two orders of magnitude smaller \citep{May}.
Since in this work we do not investigate details of the meteor
spectrum, we do not differentiate between different species of
sputtered particles. We focus only on the number of particles
originating from the surface, and therefore describe both sputtered
and scattered particles with $Y$.

The velocity of sputtered and scattered particles has one component
equal to the meteoroid velocity. An additional (smaller) component is present
because a particle picks up some additional energy from the
impact. The total velocity is a vector sum of those two components.
This makes it about 100 times faster than the average thermal speed
in the atmosphere. Such a fast particle will create a cascade of
collisions before slowing down to the thermal level. During this
process of thermalization, collisions will excite atoms and dissociate
molecules, which results in production of light by deexcitations.

We make an estimate of whether this process creates enough light to be
detected by the TV cameras used in the observations of high altitude
Leonid meteors.  During the time exposure $\Delta t$ of one video
frame, a meteor of velocity $\Vmeteor$ travels a distance $\Vmeteor
\Delta t$. The number of collisions with atmospheric particles is then
$\Rmeteor^2 \pi \Vmeteor \Delta t\, n_{\rm atm}$, where $\Rmeteor$ is
the meteoroid radius and $n_{\rm atm}$ is the atmospheric number
density. Each particle originating from the meteoroid surface will
result in many photons from excited (mostly atmospheric) atoms in the
cascade.  The total number of emitted photons is
 \eq{
     N_{\rm photon}^{tot} = \gamma Y \Rmeteor^2 \pi \Vmeteor \Delta t\, n_{\rm atm},
  }
where $\gamma$ is the total number of emitted photons in {\em one}
cascade.

These photons are emitted within the overall volume of $\sim 4\pi
(L/2)^3/3$, where high altitude meteors have size, $L$, of several
kilometers. On the other hand, one pixel in the CCD of a TV camera
covers only a fraction of the meteor angular size on the
sky. \citet{Spurny2000a} used an imaging setup with angular pixel size
$\sim$2.5''$\times$2.5''. This covers an area, $A$, of 150$\times$150
m$^2$ at 200 km distance. One pixel collects photons from the volume
$AL$.  If photons are emited uniformly within the volume of the
meteor, and asumming a meteor size of $L$=10 km, the fraction of
photons contributing to the flux measured in one pixel is
$\Psi=24A/4\pi L^2\sim$0.0004. However, photons are not emitted
uniformly, but with a radial gradient. The central region, which we
recognize as the meteor head, can emit $\sim$100 times more photons
than the average, as we will see in \S\ref{model_results} from
numerical models. This puts the estimate on $\Psi$ closer to 4\%.

If the average photon energy is $E_0$ then the measured flux in one pixel is
 \eq{
   F_{\rm m} = {\Psi E_0 N_{\rm photon}^{tot}\over 4\pi D^2 \Delta t},
 }
where $D$ is distance to the meteor. To work in units of stellar
magnitudes, we compare this flux with the flux of Sirius, $F_{\rm
sirius}$. Although the exact stellar magnitude depends on the spectral
sensitivity of the detector, we assume here a visual magnitude of
$-$1$^m$ for Sirius. The apparent meteor magnitude is then
  \eq{\label{lim_mag}
     {\mathfrak M} = -1^m - 2.5\log \left({\gamma Y \Rmeteor^2 
                                          \Vmeteor n_{\rm atm} \Psi E_0  
                                     \over 4 D^2 F_{\rm sirius}} \right). 
  }

\citet{Spurny2000a} and \citet{Fujiwara} argue that the sensitivity of
TV cameras to wavelengths of $\sim$800 nm is one of the reasons why
they can detect high altitude meteors.  Photons of that wavelength
have $E_0$=2.5$\times$10$^{-19}$ J. One collisional cascade can
produce $\gamma\sim$100 photons (see \S\ref{model_results}), as a
product of excitation by collisions or dissociation of molecules. The
highest Leonid detected by \citet{Spurny2000a} had $\Rmeteor$=0.07 m
and was first detected at distance $D\sim$200 km. By using
$\Vmeteor$=71 km/s for Leonids, $ n_{\rm atm}\sim$10$^{16}$ m$^{-3}$
and $F_{\rm sirius}\sim$2$\times$10$^{-8}$ W/m$^2$ (integrated flux
within the sensitivity range of the camera), our estimate of the
brightness is ${\mathfrak M}\sim$6.5$^m$. This is exactly the limiting
magnitude of the video system used by \citet{Spurny2000a}.

For a given observational setup, the limiting magnitude is, by
definition, the same as the meteor brightness at the beginning height,
$H_{begin}$. Equation \ref{lim_mag} shows that $H_{begin}$ depends on
the meteoroid size. Larger meteoroids reach higher maximum brightness
${\mathfrak M}^{max}$ during their disintegration below $\sim$90 km,
thus a correlation between these two parameters is to be
expected. Indeed, observations show that $H_{begin}$ is higher for
brighter meteors (figure \ref{Fig_H_mag}).  In addition to the
meteoroid size $\Rmeteor$ and distance to the beginning point $D$,
this dependence comes from variables in equation \ref{lim_mag} that
differ in two meteors from the same meteor stream.  The size of a
meteor scales with the atmospheric mean free path $l_{\rm atm}$, thus
$\Psi\propto\l_{\rm atm}^{-2}$. For altitudes between 160 km and 200
km, the atmosphere has $\l_{\rm atm}\propto\exp (H_{begin}/27.3{\rm
km})$ and $n_{\rm atm}\propto\exp (-H_{begin}/26.8{\rm km})$ (the
U.S. Standard Atmosphere 1976).

In the single body approximation of ablation, the maximum meteor brightness 
${\mathfrak M}^{max}$ in magnitudes is (see Appendix)
  \eq{
    {\mathfrak M}^{max} = -7.5\log \Rmeteor - 2.5\log (\cos Z) + {\rm constant},
  }
where $Z$ is the zenith angle of the meteoroid's trajectory.
Combining this with equation \ref{lim_mag} yields the dependence of
$H_{begin}$ on ${\mathfrak M}^{max}$:
 \eq{\label{H_red_mag}
    H_{begin} = -5.55{\rm km} \left[{\mathfrak M}^{max}\right] + {\rm constant},
  }
where $\left[{\mathfrak M}^{max}\right]$ is the {\it reduced maximum brightness}:
\eq{\label{red_mag}
   \left[{\mathfrak M}^{max}\right] = {\mathfrak M}^{max} + 7.5\log \left({D\over 200 {\rm km}} \right)
    + 2.5 \log (\cos Z)
 }
scaled to the distance of 200 km. 

Observed values of $H_{begin}$ as a function of $\left[{\mathfrak
M}^{max}\right]$ are plotted in figure \ref{Fig_H_red_mag}. Linear
regression gives a slope of -5.9$\pm$0.5 km, which agrees with the
theoretical slope in equation \ref{H_red_mag}.

\section{Monte Carlo model}

Unusual morphology is another characteristic of high altitude
meteors. In order to explain changes in their appearance with
altitude, we preformed Direct Simulation Monte Carlo (DSMC)
modeling. The simulation follows a sputtered particle and all high
velocity particles produced by collisions during the thermalization
process. The three-dimensional distribution of collisions provides
information on the spatial distribution of photons produced.

Although DSMC is widely used for modeling interactions between
rarefied gases and solids, including reentry of spacecraft
\citep{Oran}, it has very rarely been used in meteor physics.
\citet{Boyd} modeled the flow around a 1 cm Leonid meteor at 95 km
altitude. The model shows that meteoroid vapors surround the body and
interact with the atmosphere on a larger scale than when the vapors
are not produced. This confirms theoretical expectations of vapor
shielding. \citet{Zinn2004} also performed, but not with Monte Carlo,
a numerical study of Leonids at these and lower altitudes with a
hydrodynamics code coupled with radiative transfer. They found that
the meteoroid vapors are stopped very quickly by the air and moved
into the meteor wake, where the vapor-air mixture expands and
cools. This stopping process is the dominant source of energy
deposition in the atmosphere. These models demonstrate the importance
of correct numerical treatment of the vapor cloud.

All these numerical models have been applied to conditions in which
Kn$<$10, typically for altitudes below $\sim$100 km. This is where the
intense vaporization and compressed air in front of the meteoroid
create conditions appropriate for the approximation of vapor shielding
and the continuous flow regime.  This choice of flow conditions is
also influenced by the numerical codes used in simulations.  They are
not designed specifically for meteors and their applicability requires
the hydrodynamic fluid regime. In contrast, our DSMC code is written
and designed specifically for meteors under the molecular flow regime
at higher altitudes.

In this study we focus solely on the sputtering effect at altitudes
above $\sim$130 km as relevant to high altitude meteors. For
simplicity, we do not differentiate between different atomic and
molecular species. Other physical processes (like vaporization,
ionization, dissociation, excitation, chemistry, etc.) can be added in
the future to expand the applicability of the code.  The Monte Carlo
methodology used in the code is described by \citet{Xie} for the case
of cometary atmospheres. Our modifications are related to the physical
conditions in the earth's atmosphere and the sputtering process. Here
we describe the basic numerical procedure, while for the detailed
description we refer to the original paper.

The numerical simulation follows the flight of a meteoroid through the
atmosphere and calculates how many particle collisions with the air
happen during one time step $\delta t$ of calculation.  Each modeled
particle in DSMC is a statistical representative of a set of
particles.  For $N_{\rm MC}$ model particles within time step $\delta
t$ at altitude $H$, the number of real particles in one set is
 \eq{\label{Nset}
   N_{\rm set}(H) = { \Rmeteor^2 \pi \Vmeteor  n_{\rm atm}(H) \delta t(H)\, \over N_{\rm MC} }.
 }
If $\delta t$ is held constant during the simulation then the distance
covered by the meteor during one time step would eventually become
much larger than the atmospheric mean free path $\l_{\rm atm}$. In
order to avoid this statistical problem, we move the meteoroid one
$\l_{\rm atm}$ per step, thus $\delta t(H) = \l_{\rm
atm}(H)/\Vmeteor$. The number of Monte Carlo particles used in our
computations is $N_{\rm MC}$=1500.

A model particle is ejected in a random direction from a random
location on the meteoroid surface. We use a spherical meteoroid of
$\Rmeteor$=0.05 $m$ and isotropic sputtering for simplicity.  A more
general shape and sputtering angle distribution could be used in the
future.  We also assume a constant sputtering velocity $V_{\rm sp}$ of
20 km/s relative to the meteoroid.  This corresponds to $\sim$10\% of
the collision energy if the masses of projectiles and sputtered
particles are equal. We plan to introduce a more realistic
distribution of sputtered velocities, such as a Maxwellian
\citep{Coulson} or Sigmund-Thompson 
\citep{Thompson,Sigmund}.

The history of collisions of an ejected particle and all subsequent
collisionally produced fast air particles is recursively traced by the
code. The basic assumption is that fast particles collide only with
the background atmosphere, but never with each other.  This is a good
approximation as long as the local number density of ejected particles
around the meteoroid is not comparable to the atmospheric
density. This condition is satisfied in our case of sputtered
particles.

Since all modeled particles are treated as the same species, there is
only one collisional cross section $\sigma$.  We use the Variable Hard
Sphere model where the cross section is velocity dependent:
 \eq{
   \sigma =\sigma_{\rm ref}\left({ v_{\rm ref}\over v } \right)^{s}.
 }
We use the reference values typical for air: $s$=0.25, $\sigma_{\rm
ref}$=1.26$\times$10$^{-19}m^2$ and $v_{\rm ref}$=(4989/M)$^{0.5} m/s$
\citep{Boyd}, where M is the molar mass ($kg/mol$) of air at the
altitude of collision.  The time between collisions for a particle of
velocity $v$ is calculated from a random number ${\mathfrak R}$
between zero and one:
 \eq{\label{t_coll}
  t_{\rm coll} =  \sum\limits_H  \Delta t(H).
 }
 \[
  \ln ({\mathfrak R}) = -\,\,\sum\limits_H \Delta t(H) \sigma v  n_{\rm atm}(H)
 \]
This was derived from the assumption that air particles, whose
velocities are much smaller than the velocity of modeled particles,
follow the Maxwellian distribution.  The sum over $n_{\rm atm}(H)$ is
introduced because the atmospheric density is changing with altitude
along the particle trajectory.  To describe this change we use
altitude steps of 1 km. The time that the particle spends crossing an
altitude layer is $\Delta t(H)$. Since neither $t_{\rm coll}$ nor $H$
is known in advance, the equation is solved iteratively.

A collision between particles requires random selection of the
velocity vector $v_{t}$ of an air particle in the Maxwellian
distribution.  We choose cylindrical coordinates ($\varphi_t$,$v_{\rho
t}$,$v_{z t}$) where
 \eq{
  \vec{v_t} = v_{\rho t}\cos \varphi_t\, \hat{i} + 
              v_{\rho t}\sin \varphi_t\, \hat{j} + v_{z t}\,\hat{k} .
 }
From three random numbers ${\mathfrak R}_1$, ${\mathfrak R}_2$ and
${\mathfrak R}_3$, the velocity is given as:
 \eq{
    \varphi_t = 2\pi {\mathfrak R}_1
 }
 \eq{
   v_{\rho t}  = \bar{v} \sqrt{-\,\ln {\mathfrak R}_2}
 }
 \eq{
   {\rm Erf}\left({\vert v_{z t}\vert\over \bar{v}}\right) = 2 \vert{\mathfrak R}_3  - 0.5\vert 
 }
 \[
   {\rm sign}(v_{z t}) = {\rm sign}({\mathfrak R}_3  - 0.5),
 \]
where ${\rm Erf}$ is the Error integral, $\bar{v}^2$=$2RT(H)/$M$(H)$,
and $T(H)$ and M$(H)$ are the temperature and molar mass of the air at
altitude $H$. The collisional scattering of two particles is
isotropic, with velocity directions chosen randomly in the
center-of-mass frame.  When the kinetic energy of a particle drops
below some pre-defined limit (10 eV in our simulations), its
trajectory and collisions are not traced any more. This limit cannot
be below the energy of photons that we expect to be produced.  Another
reason for losing a particle is collision with the meteoroid. We do
not consider sputtering from such secondary collisions.

\section{Model results}
\label{model_results}

In order to visualize a meteor, we collect the total number of
collisions visible from the ground by a TV camera at location
$\vec{r}_{\rm cam}$. The coordinate system is shown in figure
\ref{geometry}.  The meteor's direction of flight is $\hat{v}_{\rm
m}=-\sin Z\,\hat{j}-\cos Z\,\hat{k}$, where $Z$ is the zenith angle
(45\deg in our model).  The camera's frames point at subsequent
locations along the meteor path in order to track the meteor
flight. When a collision happens at $\vec{r}_{\rm coll}$ in space and
$t_{\rm coll}$ in time (starting from the first frame), the camera's
frame number $f_{\rm cam}$ is calculated (by ignoring decimal places)
from
 \eq{
   f_{\rm cam} =  { t_{\rm coll} + \vert \vec{r}_{\rm coll} - 
                    \vec{r}_{\rm cam}\vert / c \over \Delta t }
 }
where $c$ is the speed of light.  The number of collisions in one set
$N_{\rm set}$ (equation \ref{Nset}) is added to the image pixel that
points at the collision.

The first appearance of the meteor depends on the camera's
sensitivity. The pixel values can be transformed into apparent
magnitudes if we define the frame in which the meteor first appears.
This is the frame showing the meteor at its beginning height $H_{\rm
begin}$.  The brightest pixel in that frame is set to be the limiting
magnitude ${\mathfrak M}_{\rm limit}$ and its value, $N_{\rm limit}$,
is used for calculating the apparent magnitude. If the distance of the
meteor from the camera in this first frame is $D_{\rm limit}$ then the
brightness of a pixel is
 \eq{\label{pixel_mag}
  {\mathfrak M} = {\mathfrak M}_{\rm limit} - 2.5\log \left[\left({D_{\rm limit}\over D}\right)^2 
             {N_{\rm collisions} \over N_{\rm limit}}\right],
 }
where $N_{\rm collisions}$ is the pixel value and $D$ is the meteor's distance. 

Figure \ref{meteor_models} shows images of our meteor model from four
different camera locations. The basic morphological features described
by \citet{Spurny2000b} are successfully reproduced. The meteor first
appears as a diffuse structure that grows in size, but then turns into
a comet-like V-shape, followed by a transition into a more compact,
elongated structure. The exact shape depends on the relative position
between the camera and the meteor path. Since we do not consider
photon emission from long-lived atomic states, the feature missing at
lower altitudes, below $\sim$140 km, is the train that develops behind
the meteor.  Figure \ref{meteor_models} also shows the number of
collisions ``visible'' by each pixel. Collisions are concentrated at
the center of the diffuse structure, which demonstrates the
plausibility of the value used for $\Psi$ in equation \ref{lim_mag}.

The efficiency of light production in high altitude meteors (described
as $\gamma$ in equation \ref{lim_mag}) depends on the number of
collisions.  Collisional cascades differ in their total number of
collisions. This number is mostly between 100 and 200, with $\sim$180
collisions being the most frequent (figure \ref{rate_of_coll}).  The total
energy of collision between two particles follows the distribution
$dN=Nf(E)dE$ that can be calculated from
 \eq{\label{distribution}
  f(E) = {\Delta N \over \Delta E \,\, N},
 }
where $\Delta N$ and $\Delta E$ are small increments in the number of
collisions and collisional energy, and the total number of collisions
is $N$. The distribution obtained is shown in figure
\ref{rate_of_coll_energy}. For most of its range it has the functional
dependence $f(E)\propto E^{-p}$, with $p$=0.66.

From the meteoroid's point of view, the diffuse coma does not change
much with altitude when measured in the units of atmospheric mean free
path \hbox{$l_{\rm atm}$ .}  Figure \ref{coma} shows the meteor viewed
by a camera that moves with the meteor. Since collisions between the
fast coma particles are not considered, the size of the coma is
controlled exclusively by $l_{\rm atm}$.  Decreasing altitude only
increases the number of collisions per unit volume. If the assumption
of negligible coma-coma interactions holds all the way to the end of
the molecular flow regime, then the meteor would shrink to a few
meters in size at 120 km altitude, and below one meter at $\sim$100
km. The fact that the observed meteor size is larger at these
altitudes indicates that the coma eventually becomes dense enough to
produce internal collisions. In addition, vaporization starts to
produce particles in larger numbers than sputtering, making the coma
denser than during sputtering.

Meteor images based on apparent magnitudes (upper rows in figure
\ref{meteor_models}) can be used to calculate the meteor light curve.
Total meteor brightness is the sum of all pixels brighter than the
limiting magnitude.  Hence the absolute meteor magnitude, defined as
the brightness at 100 km distance, is
 \eq{\label{meteor_mag}
  M = {\mathfrak M}_{\rm limit} - 2.5\log \left({D_{\rm limit}\over 100 {\rm km}}\right)^2 
           -2.5\log \left(\sum\limits_{m_{\rm pix}<6.5^m} 
                          {N_{\rm collisions} \over N_{\rm limit}}\right),
 }
where $m_{\rm pix}$ is the pixels' apparent magnitude.  Since $M$ is a
function of $D_{\rm limit}$, the light curve depends on the beginning
height $H_{\rm begin}$ and on the camera's position. The dependence on
$H_{\rm begin}$ is stronger, because $N_{\rm limit}$ is a function of
altitude.  Notice that according to equation \ref{lim_mag}, $N_{\rm
limit}$ is a constant for a given meteor and detector.  By letting
$N_{\rm limit}$ vary with altitude, we declare it an unknown that can
be determined experimentally from the pixel value of the limiting
magnitude in the first meteor frame.

Figure \ref{lightcurve} shows the light curves of our meteor model for
two beginning heights: 200 km and 171 km. The cameras have different
distances to the starting point of the meteor, thus they differ in the
absolute magnitude at $H_{\rm begin}$.  In the first few frames, the
meteor brightness is slightly erratic as the meteor is a few pixels in
size and susceptible to stochastic variations in pixel values.  After
that the light curve smoothes out, but it is not linear; it shows a
change in slope, which is also an observed characteristic of high
altitude meteors \citep{Spurny2000b}.

\section{Conclusion}

We have shown that thermalization of particles sputtered from a
meteoroid surface by the impinging atmosphere can explain the
phenomenon of high altitude meteors.  A sputtered particle has a much
larger kinetic energy than the surrounding atmospheric particles. It
takes between 100 and 200 collisions between particles in the
resulting collisional cascade to redistribute this initial energy down
to the thermal level of the atmosphere. The volume filled with
collisions surrounding the meteoroid is controlled by the atmospheric
mean free path.  These collisions are the source of the atomic
excitations required for the light production.

We showed that this process can create enough photons to be detected
by high-sensitivity TV cameras. The number of sputtered particles, and
consequently the number of photons produced, is proportional to the
meteoroid size.  This explains why the observed beginning height
correlates with the maximum meteor brightness. Our analytical
description of this correlation agrees with the data.  We developed a
Monte Carlo code specifically for the problem of high altitude
meteors. Simulations successfully reproduced the observed size and
morphology of these meteors. We found that their shape also depends on
the relative angle between the meteor path and the camera's line of
sight.

This model provides the grounds for further study of meteoroid
properties and meteor microphysics. Features like streams and jets
observed in high altitude meteors are probably created by nonuniform
sputtering and meteoroid rotation.  This will be addressed in the
future by Monte Carlo modeling.  It is also possible to trace
different atomic species during thermalization, which is a highly
nonequilibrium process, and to calculate their spectral lines.  Such
numerical simulations will remove the need for ``fudge'' factors
typically used to hide underlying microphysics in meteor equations.

\section*{Acknowledgment}

This work was supported by NSF grant PHY-0070928.  The author would
like to thank Peter Jenniskens, SETI Institute, and Bruce Draine,
Princeton University, for valuable discussions.

\section*{APPENDIX: Maximum brightness from the single body ablation theory}

Meteor altitude $H$ is described by the equation of motion:
 \eq{
 {dH\over dt} = - \Vmeteor \cos Z .
 }
Ablation is changing the meteoroid mass as:
 \eq{
  {dm\over dt} \propto n_{\rm atm}(H) m^{2/3} \Vmeteor^3.
 }
Combining and integrating these two equations, while keeping the meteor velocity 
constant (a good approximation for fast meteors like Leonids) and assuming 
an exponential atmosphere, we get the solution:
 \eq{
  m^{1/3}(H) = m_{\rm m}^{1/3} - {\rm constant}\times { n_{\rm atm}(H)\over \cos Z},
 }
where $m_{\rm m}$ is the initial meteoroid mass. 
At the end point of the trajectory $H_{\rm end}$, the meteor mass becomes zero, thus
 \eq{\label{n_atm}
   n_{\rm atm}(H_{\rm end}) \propto m_{\rm m}^{1/3}\cos Z.
 }
The maximum meteor light intensity is \citep[e.g.][]{Bronshten,Opik}
 \eq{\label{app_I}
  I_{\rm max} \propto m_{\rm m}^{2/3} n_{\rm atm}(H_{\rm end}).
 }
Since $m_{\rm m}^{1/3}\propto \Rmeteor$, from equation \ref{app_I} and \ref{n_atm}
it follows that the maximum brightness in magnitudes is
 \eq{
  {\mathfrak M}^{max} = -7.5\log \Rmeteor - 2.5\log (\cos Z) + {\rm constant}.
 }


\newpage
\begin{figure}
  \caption{
 Beginning heights as a function of the maximum brightness for Leonid meteors. 
 {\it Diamonds} are observations from \citet{Spurny2000a} and {\it circles} are 
 from \citet{Campbell}. Even though these two data sets are from
 different observational setups, they follow a similar functional dependence indicated
 by the {\it solid line}. Two Leonids observed by \citet{Fujiwara} are shown as 
 {\it squares}. 
  }
  \label{Fig_H_mag}
\end{figure}
\begin{figure}[h]
  \caption{
 Beginning heights as a function of the reduced maximum brightness (see equation \ref{red_mag}). 
 Only observations from \citet{Spurny2000a}, marked as {\it diamonds}, are used, in order to
 have the same limiting magnitude for all data. 
 The {\it solid line} is the linear regression, with the 95\% confidence interval shown 
 as {\it dashed lines}. For comparison, two Leonids observed by \citet{Fujiwara} are shown as 
 {\it squares}. 
  }
  \label{Fig_H_red_mag}
\end{figure}
\begin{figure}[h]
  \caption{
 Geometry of the modeled meteor and camera positions. The meteor's entry (zenith) angle
is 45\deg in the Y-Z plane and its trajectory points toward the origin of the 
coordinate system. The cameras' coordinates are: (0,150km,0) for Camera 1,
(50km,100km,0) for Camera 2, (100km,50km,0) for Camera 3 and (0,-100km,0) for Camera 4.
  }
  \label{geometry}
\end{figure}
\begin{figure}[h]
  \caption{
Modeled meteor observed by Cameras 1-4 (see figure \ref{geometry}),
respectively, from top to bottom.  The {\it upper} row in each panel
shows the meteor brightness in apparent magnitudes (see equation
\ref{pixel_mag}) based on the beginning height of 200 km and the
limiting magnitude of 6.5$^m$. The meteor's height and distance from
the camera are indicated in each frame, together with the pixel size.
The angular size of pixels is 3.75''$\times$3.75''.  The {\it lower}
row in each panel shows the number of particle collisions visible at
each pixel in these same camera frames. The meteoroid radius is
0.05 $m$.
  }
  \label{meteor_models}
\end{figure}
\begin{figure}[h]
  \caption{
 Rate of occurrence of the total number of collisions within one collisional
 cascade, calculated from 100,000 cascades.
  }
  \label{rate_of_coll}
\end{figure}
\begin{figure}[h]
  \caption{
 Distribution of collisional energies (see equation \ref{distribution}) during 
 thermalization, calculated from 1.6$\times$10$^7$ collisions. 
  }
  \label{rate_of_coll_energy}
\end{figure}
\begin{figure}[h]
  \caption{
Modeled meteor observed by a camera that follows the meteor. The pixel size,
indicated in each frame, is equal 
to the atmospheric mean free path at the meteoroid altitude. The asymmetry 
at larger altitudes comes from the variation of the mean free path in the coma.  
  }
  \label{coma}
\end{figure}
\begin{figure}[h]
  \caption{
Theoretical light curves of the high altitude meteor model for the beginning 
heights of 200 km and 171 km, observed from four different locations on the ground
(figure \ref{geometry}).
  }
  \label{lightcurve}
\end{figure}

 \noindent
 \begin{figure*}
 Fig.1:\\ \psfig{file=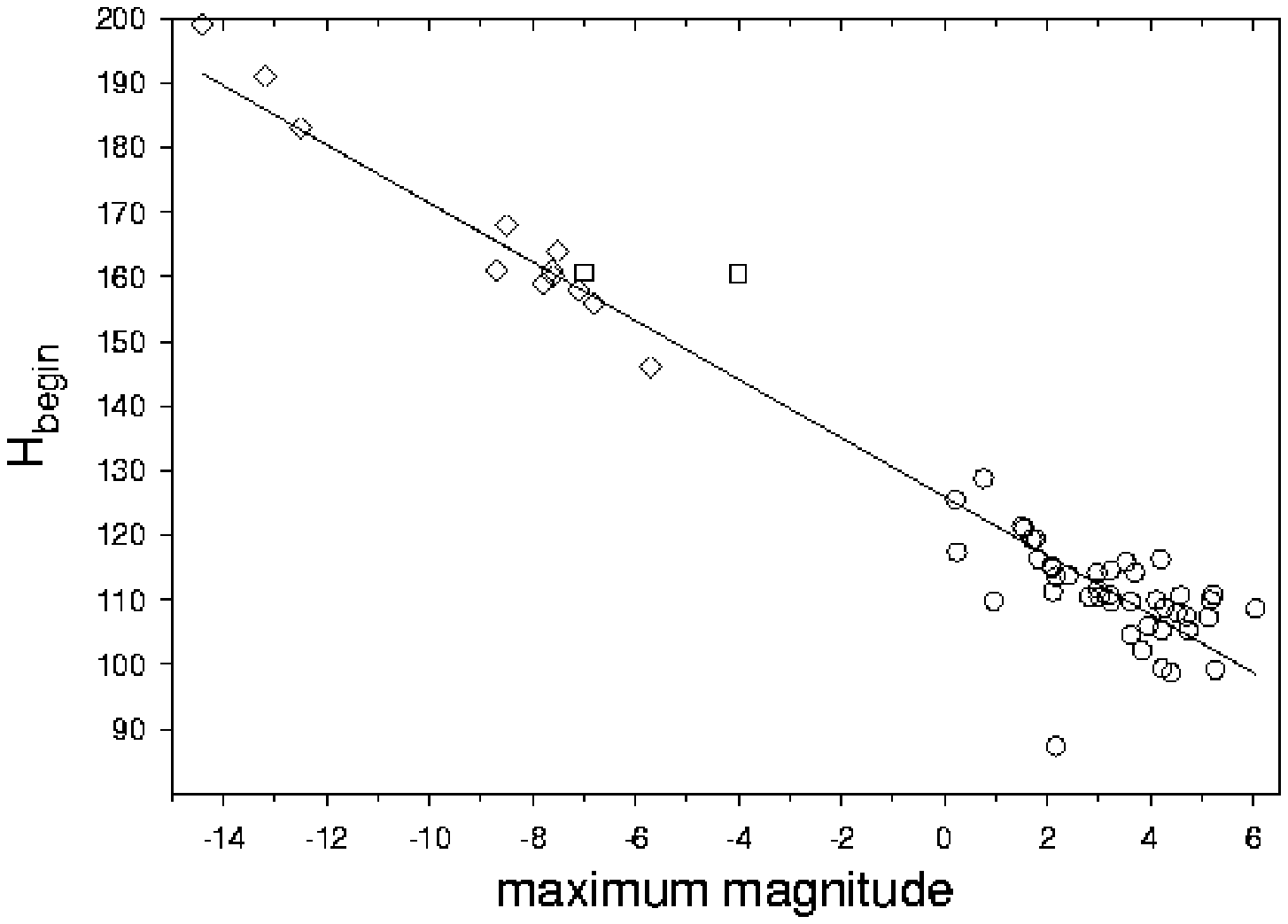,width=0.7\hsize,clip}\\
 Fig.2:\\ \psfig{file=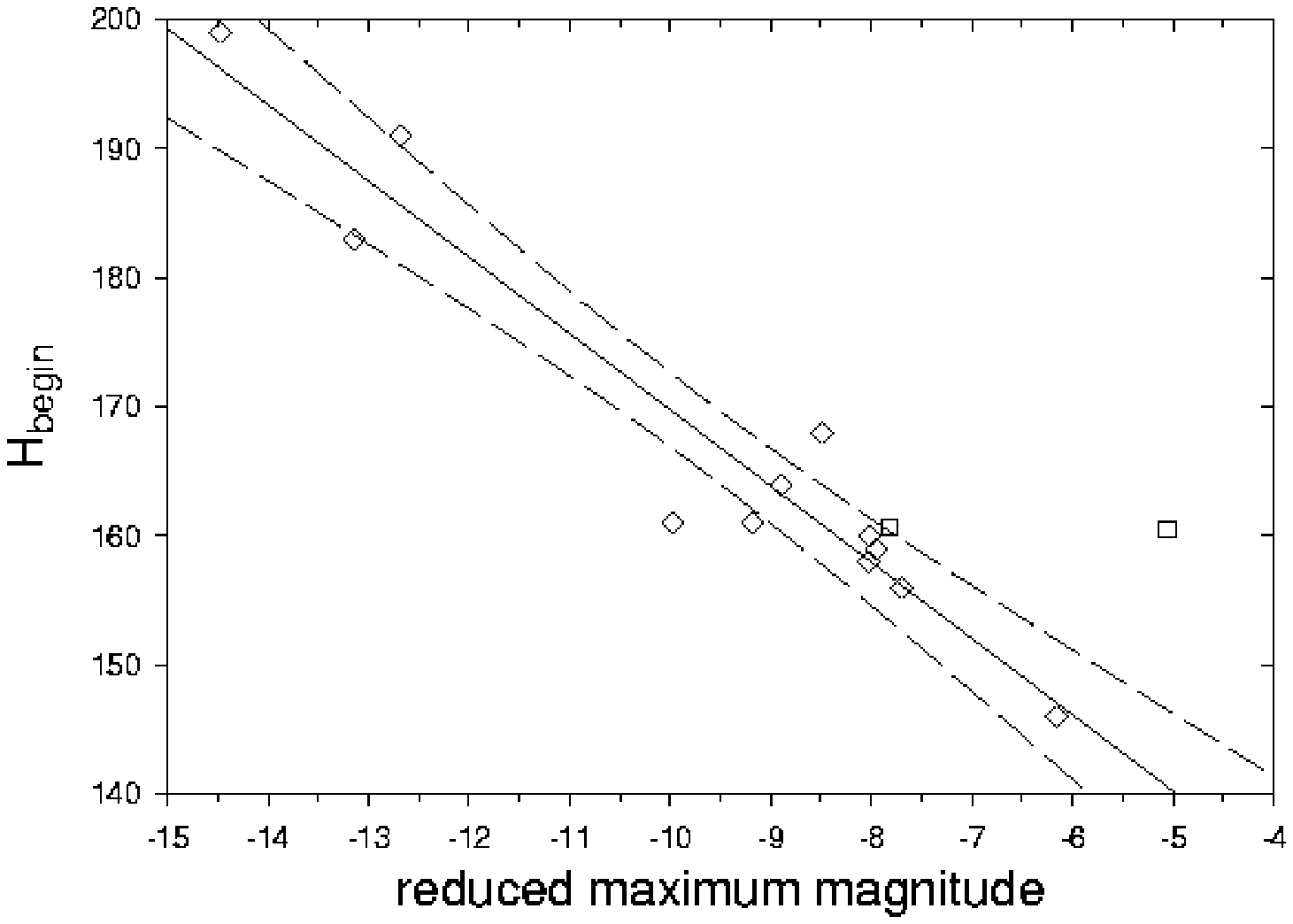,width=0.7\hsize,clip}\\
 Fig.3:\\ \psfig{file=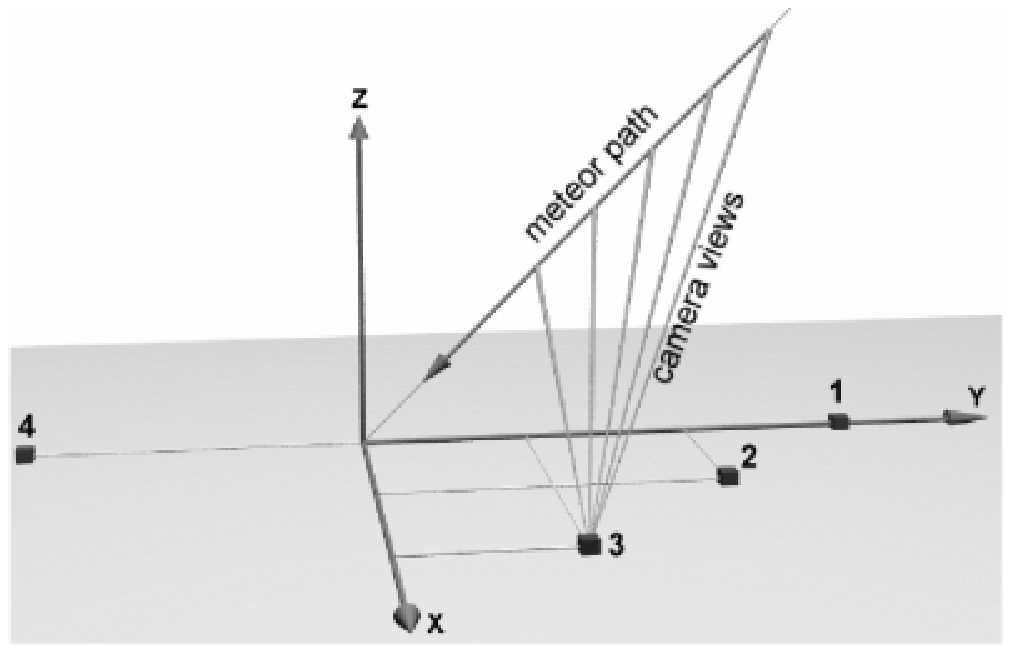,width=0.7\hsize,clip}\\
 \end{figure*}
 \begin{figure*}[h]
 Fig.4:\\ \psfig{file=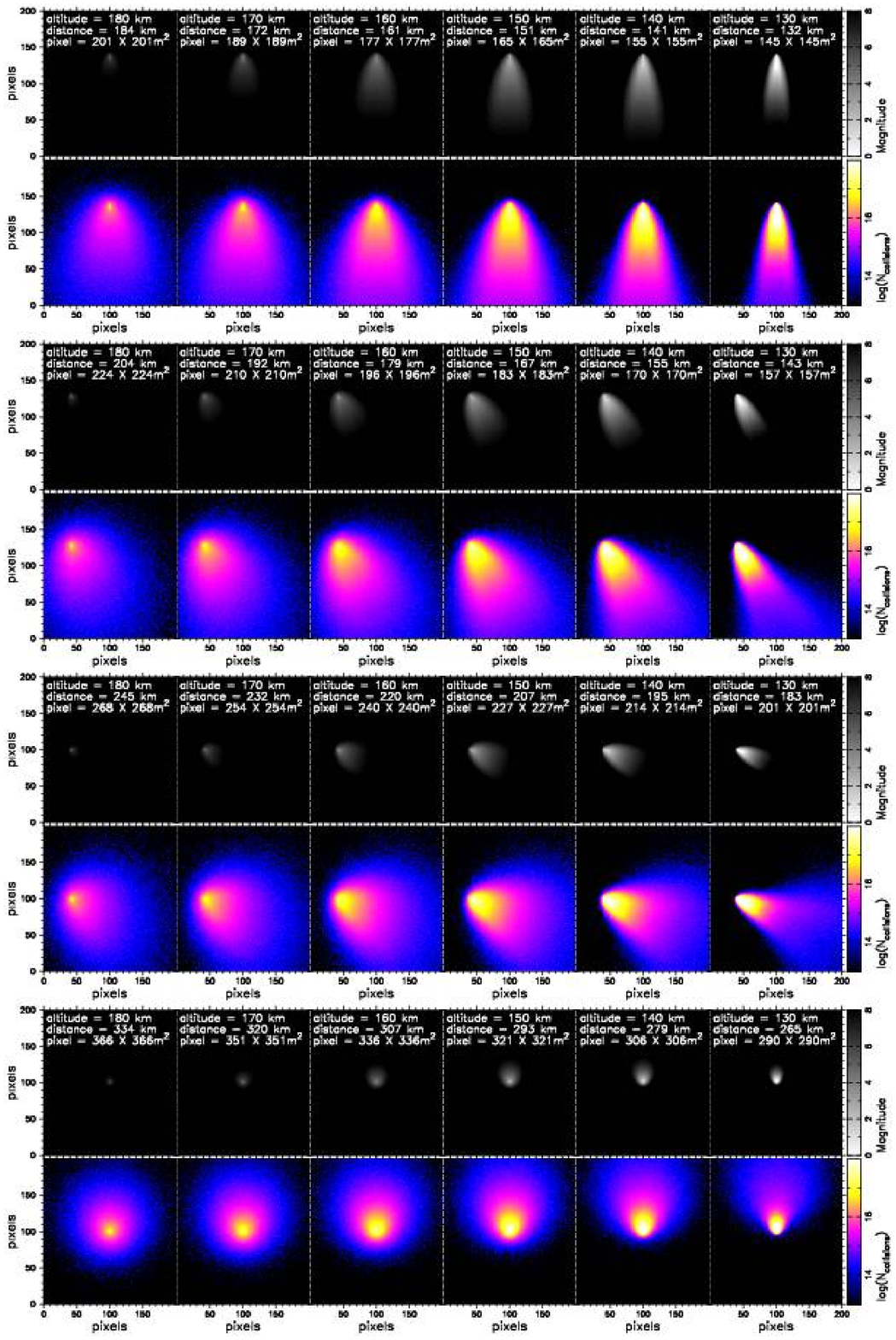,width=1.1\hsize,clip}
 \end{figure*}
 \begin{figure*}[h]
 Fig.5: \hspace{8cm} Fig.6:\\
  \psfig{file=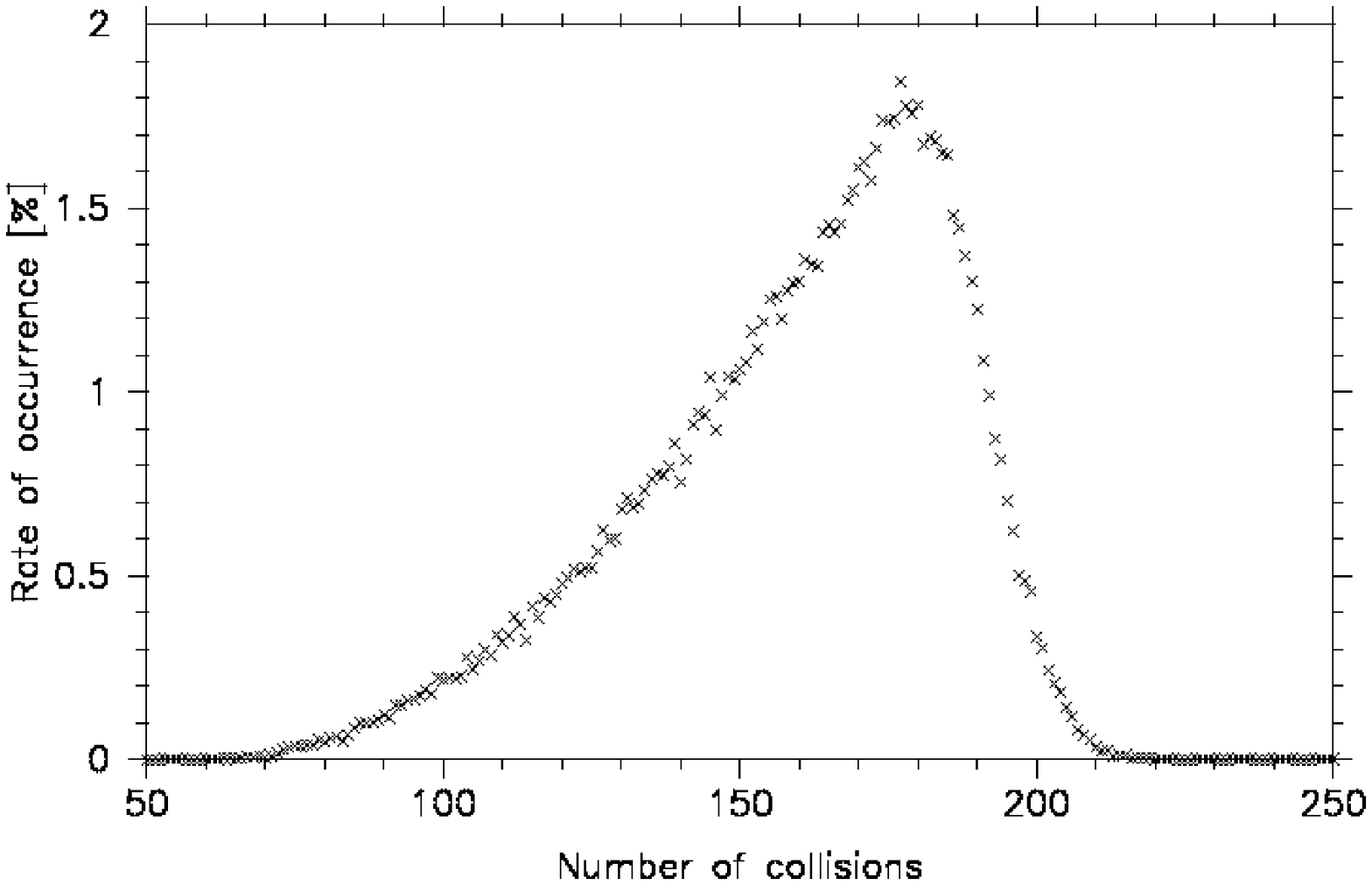,width=0.35\hsize,angle=-90,clip}
  \psfig{file=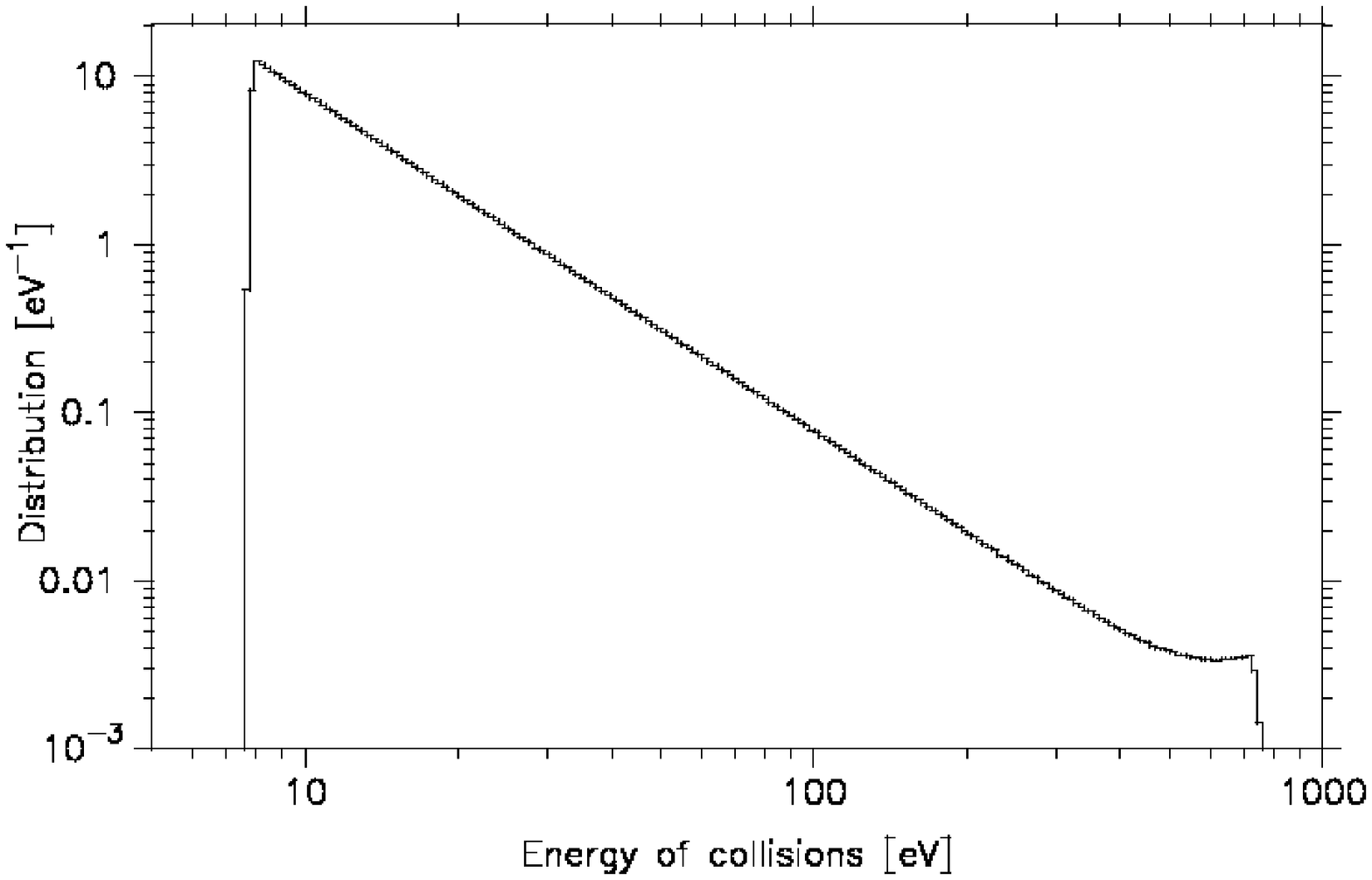,width=0.35\hsize,angle=-90,clip}\\\\
 Fig.7:\\ \psfig{file=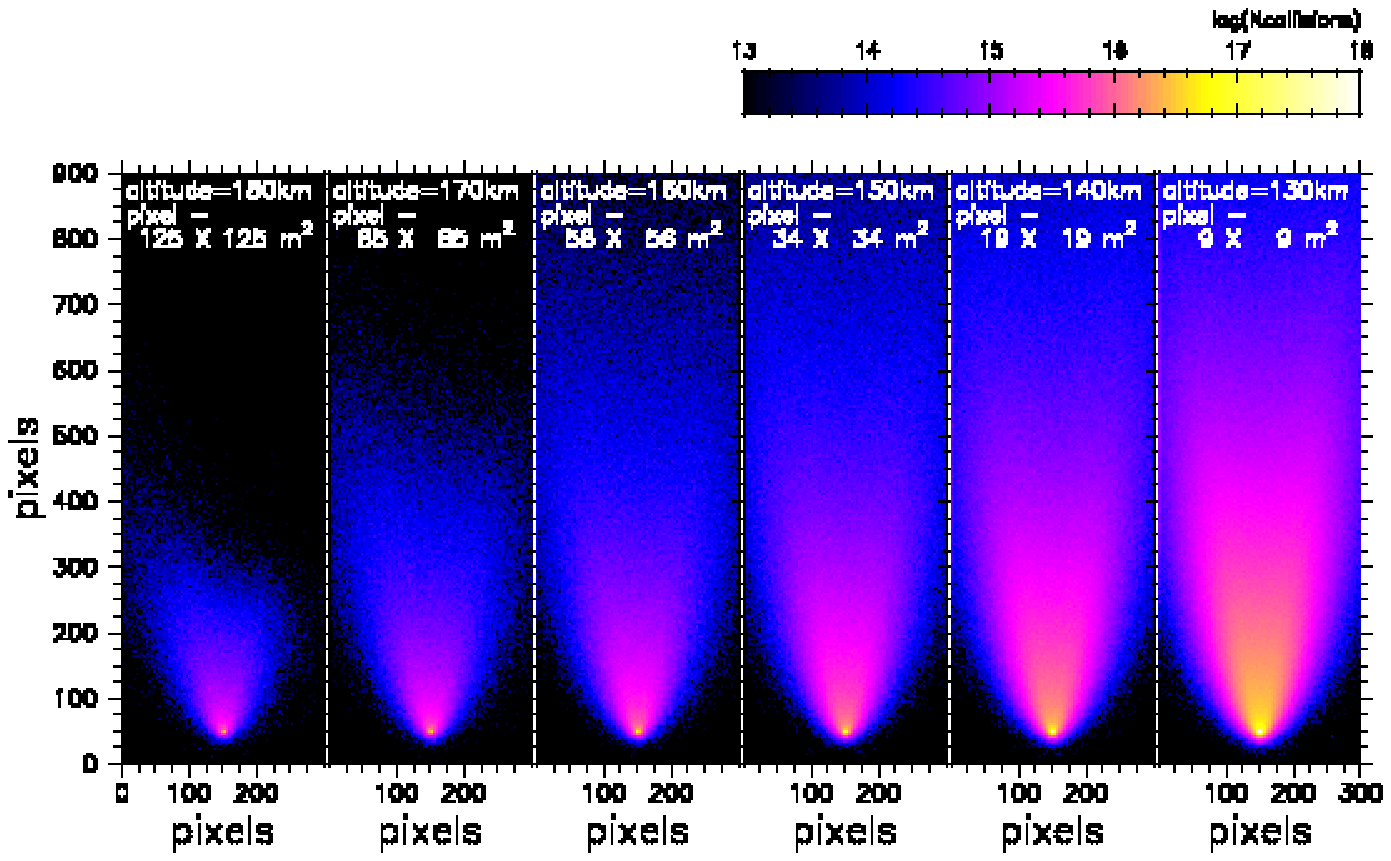,width=0.5\hsize,clip,angle=-90}\\\\
 Fig.8:\\ \psfig{file=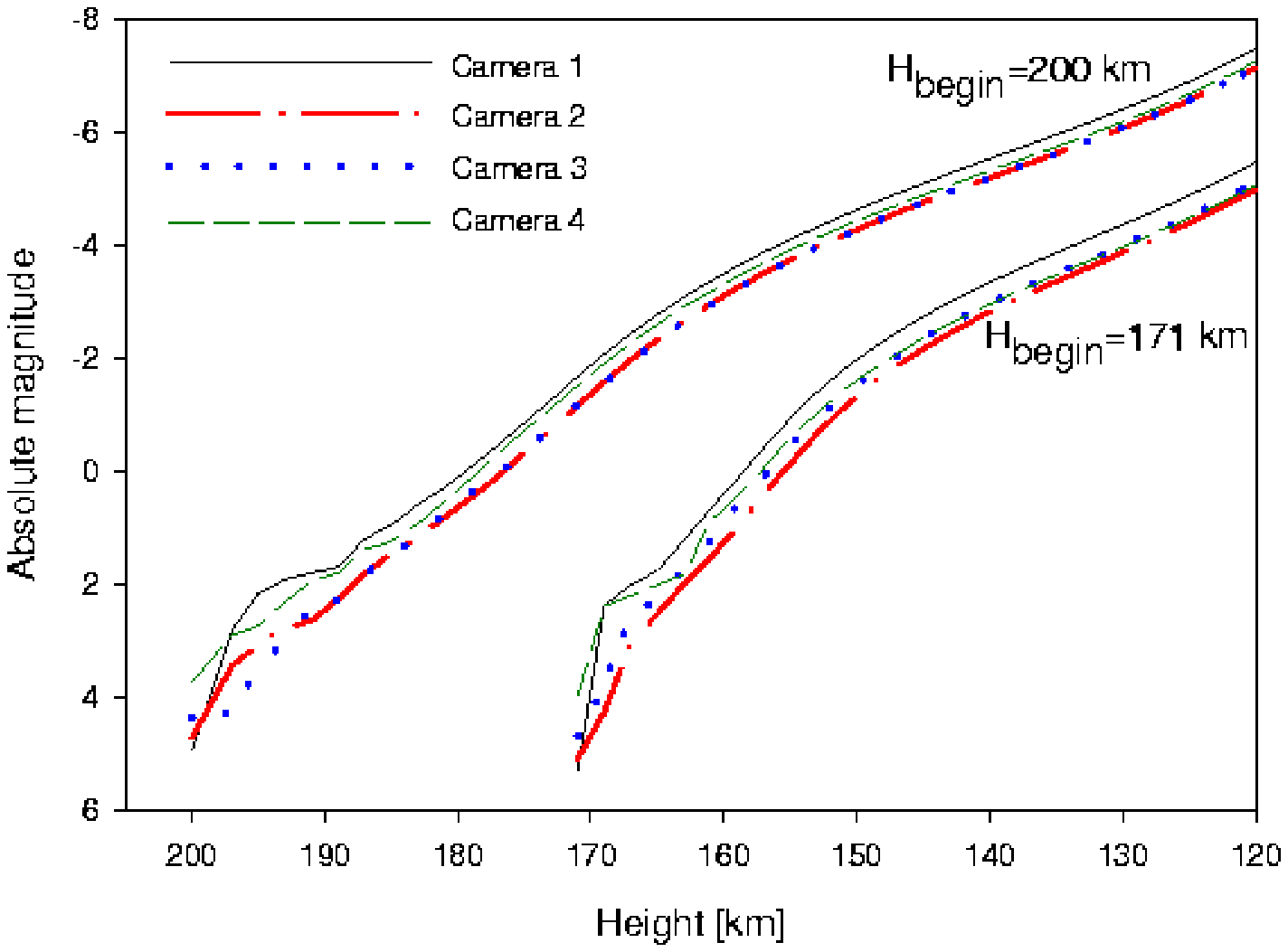,width=0.7\hsize,clip}
 \end{figure*}

\end{document}